\documentclass[aps,prx,reprint,superscriptaddress,nofootinbib,longbibliography]{revtex4-2}

\usepackage{amsmath,amssymb,mathtools,bm}
\usepackage{graphicx}
\usepackage{tikz}
\usetikzlibrary{arrows.meta,calc,positioning,decorations.pathreplacing}
\usepackage{dcolumn}
\usepackage{bbm}
\usepackage{booktabs}
\usepackage{xcolor}
\usepackage[colorlinks=true,allcolors=blue!55!black]{hyperref}
\usepackage{microtype}
\newcommand{\dd}{\mathrm{d}}
\newcommand{\ii}{\mathrm{i}}

\newcommand{\MR}{\mathrm{MR}}
\newcommand{\cH}{\mathcal{H}}

\newcommand{\cA}{\mathcal{A}}
\newcommand{\cB}{\mathcal{B}}

\newcommand{\cD}{\mathcal{D}}

\newcommand{\cQ}{\mathcal{Q}}
\newcommand{\cP}{\mathcal{P}}
\newcommand{\cS}{\mathcal{S}}
\newcommand{\cV}{\mathcal{V}}
\newcommand{\eps}{\varepsilon}
\newcommand{\Aanyon}{\mathfrak{A}}

\begin{document}

\title{An Effective String Theory Toolbox for Quantum Hall Interfaces III: \\ Open Worldsheets, Endpoint Conditions, and Branes}

\author{Ken K. W. Ma}
\affiliation{Independent researcher, Orlando, Florida 32837, USA}
\date{August 6, 2026}

\begin{abstract}
A freely moving quantum Hall (QH) interface may end on a physical edge or
topological boundary, but fixed-edge theory cannot determine what endpoint
data make such a termination consistent.  Here we formulate an open-worldsheet
junction framework in which the embedding, material charge, anomaly flow, and
topological boundary condition are organized together.  The endpoint is
specified by a geometric support and variational boundary data, together with
condensable topological sectors and any outgoing channels required to absorb
or continue the worldsheet flux.  This construction extends the charge--shape
relation to an interval and shows why a lone chiral Majorana cannot terminate
on a finite-dimensional endpoint degree of freedom.  It gives an operational
definition of a QH brane and a systematic basis for endpoint and
network theories of dynamical QH interfaces.
\end{abstract}

\maketitle

\section{Introduction}
\label{sec:introduction}

A freely moving quantum Hall (QH) interface is a chiral material boundary
whose position is itself a low-energy degree of freedom.  For a closed
interface, the companion Abelian theory relates the charged boundary sector
to the geometry of an embedded curve and derives the resulting Hall
kinematics~\cite{LiMa2021,TurkerYang2022,Ma2026I}.  A Moore--Read interface carries an additional neutral chiral Majorana sector on the same moving
worldsheet~\cite{MooreRead1991,MilovanovicRead1996,ReadGreen2000,Ma2026II}.
Those constructions apply to closed droplets or locally infinite interfaces.
They do not determine how a mobile interface may end.

The simplest open geometry is a domain wall between phases $1$ and $2$ that
terminates where both phases meet a third phase, often vacuum.  The mobile
$1|2$ segment is then connected to the $2|3$ and $3|1$ edges.  Its endpoint is
not an ordinary reflecting wall.  The first-order charged action carries a
boundary symplectic flux, chiral energy and anomaly flow must continue through
the junction, and the endpoint may absorb only those topological charges
allowed by the surrounding boundary condition.  A finite-dimensional defect
can store charge or fusion data, but it cannot by itself replace an outgoing
chiral channel.

This problem combines structures that are usually treated separately.
Open-curve mechanics determines the force, bending moment, and contact
conditions of the embedding.  The material charge--shape relation requires an
area functional completed by the support curves.  After folding, the
topological endpoint data are described by gapping or condensation data and
by any anomaly-protected channels that remain explicit.  Related ingredients
appear in Abelian Chern--Simons boundaries, anyon condensation, and defects
between gapped boundaries
~\cite{KapustinSaulina2011,BarkeshliJianQi2013,LanWangWen2015,WangWen2015,Kapustin2014,HungWan2015}.
The purpose here is to assemble them into one dynamical endpoint theory for a
moving QH interface.  This also makes precise the open-string analogy proposed
in Ref.~\cite{LiMa2021} and the endpoint information-flow questions raised in
Refs.~\cite{MaYang2022,Ma2022Scrambling}.

\begin{figure*}[t]
\centering
\includegraphics[width=0.75\textwidth]{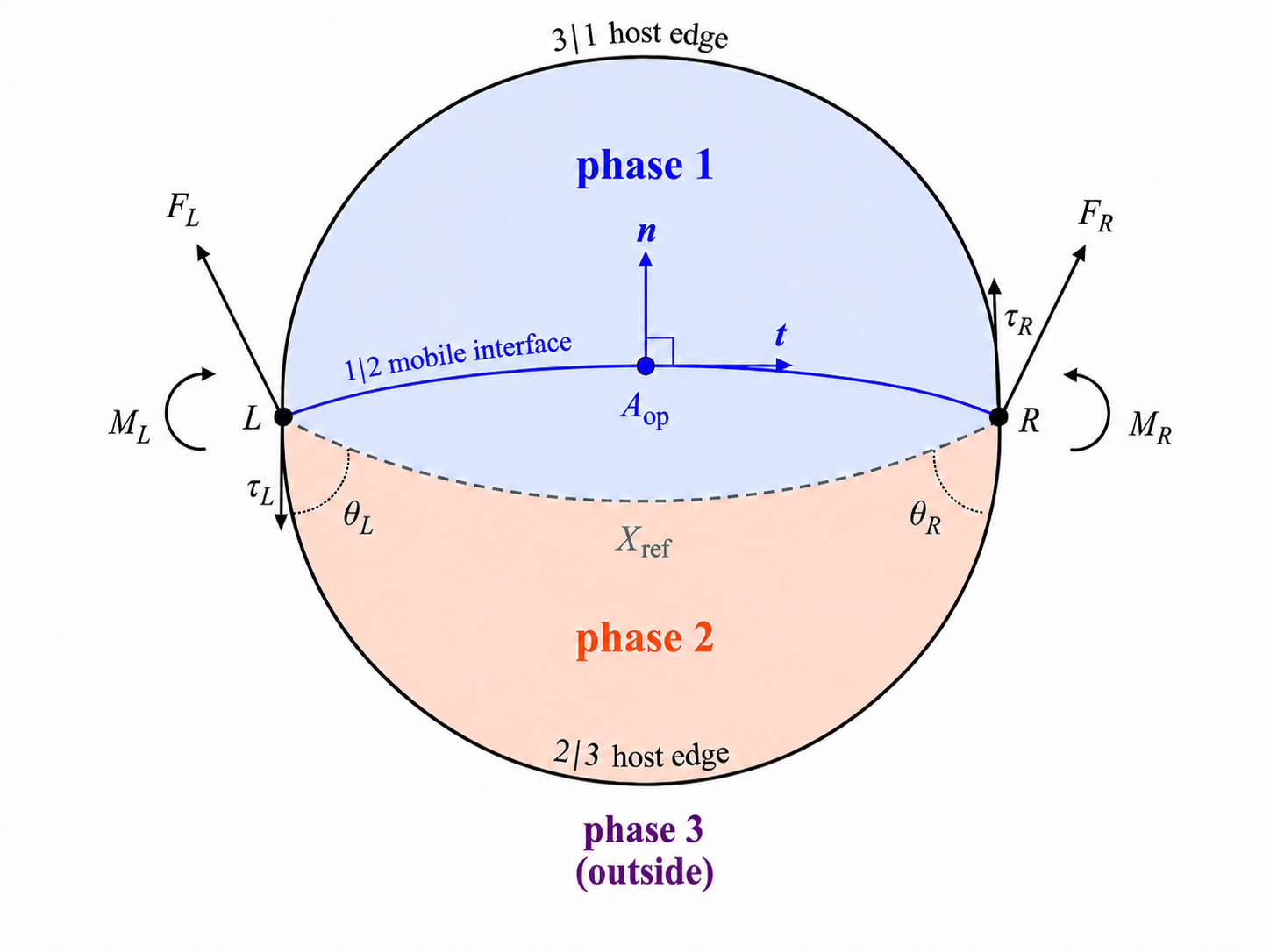}
\caption{
Schematic geometry of an open mobile QH interface.  The blue \(1|2\)
interface terminates at endpoints \(L\) and \(R\) on a circular support whose
upper and lower arcs form the \(3|1\) and \(2|3\) host edges, respectively.
The dashed curve \(\bm X_{\rm ref}\), together with the mobile segment and
the support arcs, defines the capped relative area
\(\mathcal A_{\rm op}\).  The vectors \(\bm t\) and \(\bm n\) are the unit
tangent and normal of the mobile interface.  The support tangents
\(\bm\tau_L\) and \(\bm\tau_R\) are shown with a common counterclockwise
orientation of the circular boundary, and therefore point downward at \(L\)
and upward at \(R\).  The quantities \(\bm F_a\) and \(M_a\) denote the
transmitted endpoint force and bending moment.  The dotted endpoint arcs
indicate the contact-angle convention schematically; the contact angle
\(\vartheta_a\) entering Young's law is defined by
\(\varsigma_a\bm t_a\!\cdot\!\bm\tau_a=\cos\vartheta_a\).
The displayed force and moment arrows are schematic; their signs and
directions depend on the endpoint and orientation conventions.
}
\label{fig:open_interface_geometry}
\end{figure*}

Here, we formulate an open worldsheet as a mobile embedded interval
whose endpoints lie on prescribed support curves.  The geometric variational
principle gives endpoint force and moment balance, while a capped
relative-area construction extends the material charge--shape condition to
the interval.  The no-excess material sector remains distinct from more
general endpoint charge transfer, which must be carried by an excess
interfacial sector or an explicit endpoint phase space.  At a three-phase
junction, the oriented Hall and chiral-central-charge differences telescope,
so the complete incident network has vanishing net anomaly coefficient even
though an isolated chiral segment is incomplete.  For Abelian folded sectors,
the endpoint condensates determine the topological open sectors.  For the
Moore--Read neutral sector, a local quadratic junction condition exists only
when the incoming and outgoing anomaly-carrying channels are balanced. The three-phase geometry and the endpoint notation used below are summarized
in Fig.~\ref{fig:open_interface_geometry}.

Note that our construction here is deliberately restricted to a fixed open segment and its
endpoint completion.  It does not describe interface splitting, joining, or
reconnection, and it does not assign a universal scattering matrix to a
junction.  Those amplitudes depend on microscopic endpoint physics. 

The organization of the manuscript is as follows. Section~\ref{sec:geometry} develops the open-curve geometry and mechanical boundary conditions.  Section~\ref{sec:area} constructs the capped relative area and separates material from excess charge.  Section~\ref{sec:junction} formulates the endpoint phase space and anomaly constraints.
Section~\ref{sec:abelian} describes Abelian topological boundary data and open
sectors.  Section~\ref{sec:majorana} treats Majorana junctions and Ising
boundary sectors.  Section~\ref{sec:spectrum} discusses network quantization
and endpoint response.  Technical derivations are collected in the
Appendices.

\section{Open-interface geometry and endpoint forces}
\label{sec:geometry}

\subsection{Embedded interval and support curves}

We let the mobile interface be a regular embedding of an interval,
\begin{equation}
 \bm X(\tau,\sigma),
 \qquad
 \sigma_L\leq\sigma\leq\sigma_R,
 \label{eq:open_embedding}
\end{equation}
with
\begin{align}
 \gamma&=\bm X'^2,
 &\dd s&=\sqrt\gamma\,\dd\sigma,
 \\
 \bm t&=\partial_s\bm X,
 &n^i&=-\eps^{ij}t_j.
 \label{eq:open_geometry}
\end{align}
The signed curvature and velocity components are
\begin{equation}
 K=\bm n\cdot\partial_s\bm t,
 \qquad
 \dot{\bm X}=v_t\bm t+v_n\bm n.
 \label{eq:open_K_velocity}
\end{equation}
As in the closed theory, tangential motion in the interior is a relabeling,
while normal motion changes the physical interface.

The two endpoints lie on prescribed support curves
\begin{equation}
 \bm X(\tau,\sigma_a)=\bm R_a(q_a(\tau)),
 \qquad
 a=L,R,
 \label{eq:support_constraint}
\end{equation}
where $q_a$ is a coordinate along the support and
\begin{equation}
 \bm\tau_a
 =\frac{\partial_{q_a}\bm R_a}{|\partial_{q_a}\bm R_a|}
 \label{eq:support_tangent}
\end{equation}
is its unit tangent.  The support may be a pinned QH edge, a gapped boundary,
or a narrow physical channel.  Endpoint variations are restricted to
\begin{equation}
 \delta\bm X_a
 =\bm\tau_a\,\delta\ell_a,
 \label{eq:endpoint_allowed_variation}
\end{equation}
where $\delta\ell_a=|\bm R_a'|\delta q_a$.  The tangent of the mobile interface
is otherwise allowed to rotate, unless an endpoint torque fixes the contact
angle.

\subsection{Boundary variation of tension and bending energy}

Consider the tension-plus-bending energy
\begin{equation}
 E_{0+2}[\bm X]
 =T_0\int\dd s
 +T_2\int\dd s\,K^2,
 \label{eq:open_E02}
\end{equation}
using the same curvature convention as in the closed theory
~\cite{LangerSinger1984,Ma2026I}.
For a general variation
\begin{equation}
 \delta\bm X=\eta_t\bm t+\eta_n\bm n,
 \label{eq:open_variation}
\end{equation}
we use
\begin{align}
 \delta(\dd s)
 &=(\partial_s\eta_t-K\eta_n)\dd s,
 \\
 \delta K
 &=\partial_s^2\eta_n+K^2\eta_n+\eta_t\partial_sK.
 \label{eq:open_variation_identities}
\end{align}
Integration by parts gives
\begin{widetext}
\begin{equation}
 \delta E_{0+2}
 =\int\dd s\,
 \left[-T_0K+T_2(2\partial_s^2K+K^3)\right]\eta_n
 +\left[
 (T_0+T_2K^2)\eta_t
 +2T_2K\partial_s\eta_n
 -2T_2(\partial_sK)\eta_n
 \right]_{L}^{R}.
 \label{eq:open_energy_variation_raw}
\end{equation}
\end{widetext}
The endpoint tangent angle varies as
\begin{equation}
 \delta\theta
 =\partial_s\eta_n+K\eta_t.
 \label{eq:delta_theta}
\end{equation}
Hence the boundary term takes the mechanical form
\begin{equation}
 \delta E_{0+2}\big|_{\partial\Sigma}
 =\left[
 \bm F\cdot\delta\bm X
 +M\,\delta\theta
 \right]_{L}^{R},
 \label{eq:force_moment_form}
\end{equation}
where
\begin{equation}
 \bm F
 =(T_0-T_2K^2)\bm t
 -2T_2(\partial_sK)\bm n,
 \quad
 M=2T_2K.
 \label{eq:force_and_moment}
\end{equation}
The bulk normal force agrees with the closed-curve result.  The new data are
the transmitted endpoint force $\bm F$ and bending moment $M$.  The
integration by parts and the conversion to the force--moment form are given in
Appendix~\ref{app:geometric_variation}.

Let the endpoint carry a local energy $U_a(q_a,\theta_a)$ and define
\begin{equation}
 \varsigma_L=-1,
 \qquad
 \varsigma_R=+1.
 \label{eq:end_signs}
\end{equation}
Stationarity under endpoint translation and rotation gives
\begin{align}
 \varsigma_a\bm F_a\cdot\bm\tau_a
 +\partial_{\ell_a}U_a
 &=0,
 \label{eq:end_force_balance}
 \\
 \varsigma_a M_a
 +\partial_{\theta_a}U_a
 &=0.
 \label{eq:end_moment_balance}
\end{align}
Eq.~\eqref{eq:end_moment_balance} gives $K_a=0$ for a freely hinged end
when $U_a$ is angle independent.  A clamped endpoint instead fixes
$\theta_a$ and does not impose that condition.

For $T_2=0$, let motion of the contact point replace a length of phase-$1$
edge by phase-$2$ edge, so that
\begin{equation}
 U_a=(\tau_{1a}-\tau_{2a})\ell_a.
 \label{eq:wall_tensions}
\end{equation}
Then Eq.~\eqref{eq:end_force_balance} becomes the Young relation
\begin{equation}
 T_0\cos\vartheta_a
 =\tau_{2a}-\tau_{1a},
 \label{eq:Young_law}
\end{equation}
with $\vartheta_a$ defined using the orientation for which
$\varsigma_a\bm t_a\cdot\bm\tau_a=\cos\vartheta_a$.  Thus the geometric part of an open QH worldsheet obeys the same static
contact-angle law as an ordinary capillary interface
~\cite{deGennes2004}, while its time evolution remains Hall chiral rather than
inertial or dissipative.

\section{Open relative area and the material charge constraint}
\label{sec:area}

\subsection{Capped relative-area functional}

For a closed loop, the material charge is tied to the enclosed area.  For an
open curve, an area is defined only after the mobile segment is capped by the
support curves.  Choose fixed reference points on the two supports and define
one-dimensional primitives
\begin{equation}
 \frac{\dd\cB_a}{\dd q_a}
 =\frac12\eps_{ij}R_a^i(q_a)\frac{\dd R_a^j}{\dd q_a}.
 \label{eq:brane_area_primitive}
\end{equation}
The primitive $\mathcal B_a$ is local in the support coordinate.  If the
support is a closed curve, it is defined patchwise; changing its branch shifts
$\mathcal A_{\rm ref}$ by a constant and does not affect the variation or the
material charge relation below. The open relative area is
\begin{equation}
\begin{aligned}
 \cA_{\rm op}[\bm X,q_L,q_R]
 ={}&-\frac12
 \int_{\sigma_L}^{\sigma_R}\dd\sigma\,
 \eps_{ij}X^iX'^{j}
 \\
 &+\cB_R(q_R)-\cB_L(q_L)-\cA_{\rm ref}.
\end{aligned}
 \label{eq:open_relative_area}
\end{equation}
The constant $\cA_{\rm ref}$ fixes the reference configuration.  A direct
variation gives
\begin{equation}
 \delta\cA_{\rm op}
 =\int\dd s\,\eta_n.
 \label{eq:open_area_variation}
\end{equation}
The endpoint terms from the open line integral cancel exactly against the
variations of $\cB_L$ and $\cB_R$.  Therefore, for stationary support curves,
\begin{equation}
 \frac{\dd\cA_{\rm op}}{\dd\tau}
 =\int\dd s\,v_n.
 \label{eq:open_area_time}
\end{equation}
Sliding an endpoint tangentially along its support produces no separate
endpoint term in the area variation; any physical change is already contained
in the normal deformation of the mobile curve.  The cancellation is verified
in Appendix~\ref{app:open_area}.

Eq.~\eqref{eq:open_relative_area} is the two-dimensional analogue of the
boundary completion of an open Wess--Zumino term
~\cite{FigueroaOFarrillMohammedi2005}.  A change of cap or reference shifts
$\cB_a$ and $\cA_{\rm ref}$ together and leaves all variations unchanged.  The
support curve therefore carries a geometric connection even before
topological endpoint labels are added.

\subsection{Local transgression and equal-time constraint}

The relative-area connection of Ref.~\cite{Ma2026I} extends to an interval.
Let $\bm Y(\tau,\sigma,r)$ interpolate between a reference interface and the
physical interface, while its endpoints remain on the corresponding support
curves for every $r\in[0,1]$.  Define
\begin{equation}
 \mathfrak a_a
 =-B\int_0^1\dd r\,
 \eps_{ij}\partial_rY^i\partial_aY^j,
 \qquad a=\tau,\sigma.
 \label{eq:open_transgression}
\end{equation}
The closedness of the magnetic area form gives
\begin{equation}
 \partial_\tau\mathfrak a_\sigma
 -\partial_\sigma\mathfrak a_\tau
 =B\sqrt\gamma\,v_n.
 \label{eq:open_transgression_curvature}
\end{equation}
Its integral reproduces the capped relative area,
\begin{equation}
 \int_{\sigma_L}^{\sigma_R}\dd\sigma\,
 \mathfrak a_\sigma
 =B\cA_{\rm op},
 \label{eq:open_transgression_integral}
\end{equation}
up to the fixed reference convention.  Because both
$\partial_r\bm Y$ and $\partial_\tau\bm Y$ are tangent to the same
one-dimensional support at an endpoint,
\begin{equation}
 \mathfrak a_\tau(\sigma_L)
 =\mathfrak a_\tau(\sigma_R)=0.
 \label{eq:open_transgression_endpoint}
\end{equation}
The integrated form of Eq.~\eqref{eq:open_transgression_curvature} is therefore
consistent with Eq.~\eqref{eq:open_area_time}.  The transgression and endpoint
identities are derived in Appendix~\ref{app:open_area}.

For an interface between Abelian phases, define
\begin{equation}
 \Delta\nu=\nu_1-\nu_2,
 \qquad
 \Delta\rho=\frac{B\Delta\nu}{2\pi}.
 \label{eq:open_deltanu}
\end{equation}
In the sector with no independently stored interfacial charge, the charged
field $\varphi_c$ obeys the material condition
\begin{equation}
 \cQ_{\rm mat}(\sigma)
 \equiv
 \varphi_c'(\sigma)
 -\Delta\nu\,\mathfrak a_\sigma(\sigma)
 =0.
 \label{eq:open_material_constraint}
\end{equation}
Integrating over the interval gives
\begin{equation}
 Q_{\rm mat}
 \equiv
 \frac{\varphi_c(\sigma_R)-\varphi_c(\sigma_L)}{2\pi}
 =\Delta\rho\,\cA_{\rm op}.
 \label{eq:open_total_charge_area}
\end{equation}
Thus the difference of the endpoint bosons measures the material charge
associated with the displaced bulk fluids.  At a physical endpoint these
fields must be glued to the charge variables of the host edges or to an
explicit endpoint phase space.

Generic endpoint charge transfer belongs to a larger sector.  Define the
excess density and its integral by
\begin{align}
 q_{\rm ex}(\sigma)
 &=
 \frac{1}{2\pi}
 \left(
 \varphi_c'-\Delta\nu\,\mathfrak a_\sigma
 \right),
 \label{eq:open_excess_density}
 \\
 Q_{\rm ex}
 &=
 \int_{\sigma_L}^{\sigma_R}\dd\sigma\,q_{\rm ex}.
 \label{eq:open_excess_charge}
\end{align}
The total charged-boson zero mode on the segment is then
\begin{equation}
 Q_{\rm seg}
 =
 \frac{\varphi_c(\sigma_R)-\varphi_c(\sigma_L)}{2\pi}
 =
 \Delta\rho\,\cA_{\rm op}+Q_{\rm ex}.
 \label{eq:open_charge_decomposition}
\end{equation}
If $j_L$ and $j_R$ denote excess currents entering at the left endpoint and
leaving at the right, charge conservation gives
\begin{align}
 \frac{\dd Q_{\rm ex}}{\dd\tau}
 &=j_L-j_R,
 \label{eq:open_excess_balance}
 \\
 \frac{\dd Q_{\rm seg}}{\dd\tau}
 &=j_L-j_R
 +\Delta\rho\int\dd s\,v_n.
 \label{eq:segment_charge_balance}
\end{align}
The strict material sector is $q_{\rm ex}=0$ and contains no independent
endpoint transfer into the mobile segment.  A junction that stores or
transmits additional charge must retain $Q_{\rm ex}$ or an explicit endpoint
charge variable.  This distinction is the open-interface analogue of the
material and excess-charge sectors in the closed theory.

\section{Endpoint completion as a junction theory}
\label{sec:junction}

\subsection{Why an isolated chiral segment is incomplete}

To expose the endpoint symplectic flux, for $\Delta\nu\neq0$ consider the folded charged action on a prescribed interval, or equivalently its linear material reduction,
\begin{equation}
 S_c
 =-\frac{1}{4\pi\Delta\nu}
 \int\dd\tau\int_{\sigma_L}^{\sigma_R}\dd\sigma\,
 \varphi_c'\dot\varphi_c
 -\int\dd\tau\,H_c.
 \label{eq:charged_interval_action}
\end{equation}
This equation is used only to identify the boundary flux.  It is not a claim
for a complete nonlinear first-order action of the moving embedding, whose
limitations were discussed in Paper I~\cite{Ma2026I}.
Its variation contains the endpoint symplectic flux
\begin{equation}
 \delta S_c\big|_{\partial\Sigma}
 =-\frac{1}{4\pi\Delta\nu}
 \int\dd\tau\,
 \left[\delta\varphi_c\,\dot\varphi_c\right]_L^R
 +\delta S_H\big|_{\partial\Sigma}.
 \label{eq:charged_endpoint_flux}
\end{equation}
Fixing $\delta\varphi_c=0$ would eliminate endpoint charge transfer and is not
the generic physical situation.  A consistent open theory must cancel this
flux against the variations of the host-edge actions or an endpoint phase
space.  The same conclusion follows from anomaly inflow: a chiral current algebra
cannot simply stop at a point~\cite{CallanHarvey1985,GromovJensenAbanov2016}.

Suppose phases $1$, $2$, and $3$ meet at a trijunction.  Orient the three
incident interfaces cyclically.  Their electromagnetic anomaly coefficients
are
\begin{equation}
 \Delta\nu_{ij}=\nu_i-\nu_j.
 \label{eq:interface_anomaly_coeff}
\end{equation}
They satisfy the exact telescoping identity
\begin{equation}
 \Delta\nu_{12}+\Delta\nu_{23}+\Delta\nu_{31}=0.
 \label{eq:hall_telescoping}
\end{equation}
Likewise, if $c_i$ is the chiral central charge of phase $i$,
\begin{equation}
 \Delta c_{12}+\Delta c_{23}+\Delta c_{31}=0.
 \label{eq:central_telescoping}
\end{equation}
For a mobile $1|2$ interface ending on the physical edge, one may take phase
$3$ to be vacuum.  The two pinned host edges then provide the additional channels through
which the Hall and gravitational anomaly flow may continue.  The telescoping
identities ensure that the complete three-interface network has no net bulk
response mismatch at the junction.  They are necessary but not sufficient:
they do not determine the allowed local operators, the endpoint Hilbert space,
or the microscopic scattering matrix
~\cite{CallanHarvey1985,GromovJensenAbanov2016,Stone2012}.

\subsection{Local endpoint phase space and current conservation}

If an endpoint can store electric charge, introduce a compact phase
$\chi_a$ and charge $Q_a$ with action
\begin{equation}
 S_{Q,a}
 =\int\dd\tau\,
 \left[
 Q_a\dot\chi_a-H_a(Q_a,q_a)
 \right].
 \label{eq:endpoint_charge_action}
\end{equation}
The canonical bracket is
\begin{equation}
 \{\chi_a,Q_b\}=\delta_{ab}.
 \label{eq:endpoint_charge_bracket}
\end{equation}
The endpoint gluing constraint identifies $\chi_a$ with an allowed
integer combination of the incident edge bosons.  Its equation of motion is
the junction continuity law
\begin{equation}
 \dot Q_a+\sum_{\alpha\in J_a}s_{\alpha a}j_\alpha(0)=0,
 \label{eq:junction_continuity}
\end{equation}
where $s_{\alpha a}=+1$ for a current flowing out of the junction and $-1$
for a current flowing in.  For the mobile segment, the swept-area contribution and the excess
endpoint current are separated by Eqs.~\eqref{eq:open_charge_decomposition}
and~\eqref{eq:segment_charge_balance}.
When no low-energy endpoint charge is allowed, $Q_a$ is constrained to a fixed
topological value and Eq.~\eqref{eq:junction_continuity} reduces to current
conservation among the incident channels.

For several Abelian edge fields, collect all incident modes after folding
into one vector $\bm\Phi_J$.  A useful class of local strong-coupling boundary
conditions is generated by
\begin{equation}
 U_J
 =-\sum_{p}g_p
 \cos(\bm\ell_p^T\bm\Phi_J-\alpha_p).
 \label{eq:junction_cosines}
\end{equation}
A mutually commuting, charge-conserving set satisfies
\begin{align}
 \bm\ell_p^T K_J^{-1}\bm\ell_q&=0,
 \label{eq:junction_null}
 \\
 \bm t_J^T K_J^{-1}\bm\ell_p&=0.
 \label{eq:junction_charge_neutral}
\end{align}
Each $\bm\ell_p$ must also represent a primitive microscopically local
tunneling operator.  When the perturbations are relevant and a sufficient
number of independent vectors is present, they pin the nonanomalous
combinations.  Any unpaired charged or neutral mode remains as a physical
outgoing channel.  These conditions give a sufficient Abelian realization of
the statement that a brane absorbs only mutually condensable,
charge-neutral topological excitations
~\cite{Haldane1995,KapustinSaulina2011,SantosHughes2017,MayMannHughes2019}.

\subsection{Operational definition of a brane}

For the present worldsheet theory, the minimal brane data are the
quadruple
\begin{equation}
 \mathfrak B_a
 =\left(
 \Gamma_a,
 U_a,
 \Lambda_a,
 \cH_a
 \right).
 \label{eq:brane_definition}
\end{equation}
Here $\Gamma_a$ is the support geometry, $U_a$ determines the mechanical
contact conditions, and $\Lambda_a$ denotes the condensation or gluing data;
in the folded Abelian $K$-matrix setting it is an isotropic lattice.
The factor $\cH_a$ is the endpoint or outgoing-channel Hilbert space,
together with the class of allowed local couplings to the incident worldsheet
modes.  Two
supports with the same geometry but different $\Lambda_a$ are physically
different branes.  Conversely, a topological label without
Eqs.~\eqref{eq:end_force_balance}, \eqref{eq:end_moment_balance}, and
\eqref{eq:junction_continuity} is not yet a dynamical endpoint theory.

This definition is intentionally nonrelativistic.  There is no Weyl symmetry,
no requirement of Neumann or Dirichlet conditions in every target-space
direction, and no fundamental Chan--Paton gauge group.  The string-theory
analogy is useful because the relative-area term requires endpoint completion
and because open sectors depend on labels at both ends.  The physical content
is entirely QH: support curves, Hall anomaly flow, anyon condensation, and
fusion spaces.

\section{Abelian branes and open sectors}
\label{sec:abelian}

\subsection{Lagrangian subgroups and endpoint absorption}

Let a nonchiral folded Abelian topological order have anyon group
\begin{equation}
 \Aanyon
 =\mathbb Z^N/K\mathbb Z^N.
 \label{eq:anyon_group}
\end{equation}
The braiding phase of representatives $\bm a,\bm b\in\mathbb Z^N$ is
\begin{equation}
 M_{ab}
 =\exp\left(
 2\pi\ii\,\bm a^TK^{-1}\bm b
 \right).
 \label{eq:abelian_monodromy}
\end{equation}
Within the bosonic Abelian $K$-matrix setting considered here, a fully
gapped boundary is specified by a Lagrangian subgroup
$L\subset\Aanyon$: its elements have trivial
topological spin and mutual braiding, and every anyon outside $L$ braids
nontrivially with at least one element of $L$
~\cite{BaisSlingerland2009,KapustinSaulina2011,KitaevKong2012,Levin2013,BarkeshliJianQi2013}.
Anyons in $L$ can end on the boundary without leaving a topological
excitation.  For an electronic or more general fermionic topological order,
the corresponding statement is formulated using the appropriate fermionic
condensable algebra after local electrons have been identified~\cite{WanWang2017,AasenLakeWalker2019}.  The ordinary
Lagrangian-subgroup formulas below apply to the resulting Abelian bosonic
quotient.

The construction applies directly to a gappable folded interface or to a
nonchiral collection of channels at a junction.  A strictly chiral QH edge
with nonzero net chiral central charge cannot be fully gapped by itself.  In
that case $L$ labels only the gappable subsector, while the anomaly-mandated
outgoing channels remain explicit.

\subsection{Open Wilson-line sectors}

Consider a strip with branes $A$ and $B$ characterized by Lagrangian
subgroups $L_A$ and $L_B$.  A Wilson line carrying anyon $a$ may stretch
between the two ends.  Its label is unchanged if one attaches a condensed
anyon $\ell_A\in L_A$ at the left endpoint or $\ell_B\in L_B$ at the right.
Therefore
\begin{equation}
 a\sim a+\ell_A+\ell_B.
 \label{eq:anyon_equivalence}
\end{equation}
The set of topological open sectors is the double quotient
\begin{equation}
 \cS_{AB}
 =L_A\backslash\Aanyon/L_B
 \simeq\Aanyon/(L_A+L_B),
 \label{eq:open_sector_quotient}
\end{equation}
where the second form uses the Abelian group law.  The topological Hilbert
space is the complex vector space with this sector basis,
\begin{equation}
 \cH_{AB}^{\rm top}
 \simeq\mathbb C[\cS_{AB}].
 \label{eq:open_sector_Hilbert}
\end{equation}
When both subgroups are Lagrangian,
\begin{equation}
 |L_A|=|L_B|=\sqrt{|\Aanyon|}.
 \label{eq:Lagrangian_size}
\end{equation}
Using
\begin{equation}
 |L_A+L_B|
 =\frac{|L_A||L_B|}{|L_A\cap L_B|},
 \label{eq:subgroup_sum_size}
\end{equation}
we obtain
\begin{equation}
 \dim\cH_{AB}^{\rm top}
 =|\cS_{AB}|
 =|L_A\cap L_B|.
 \label{eq:open_sector_dimension}
\end{equation}
Gapless oscillator modes and endpoint charging states multiply this
topological sector.  The counting is derived in
Appendix~\ref{app:open_counting} and agrees with the boundary-dependent
ground-state degeneracy of Abelian topological orders on a cylinder or open surface~\cite{LanWangWen2015,WangWen2015,Kapustin2014,HungWan2015}.

The ordered pair $(A,B)$ plays the endpoint-label role analogous to
Chan--Paton data~\cite{Paton1969}, while $[a]\in L_A\backslash\mathfrak A/L_B$ labels the allowed topological Wilson-line sector within the corresponding $AB$ block.  No independent matrix multiplicity or Chan--Paton gauge group is implied.

\subsection{$\mathbb Z_N$ example and endpoint zero modes}

For the doubled $\mathbb Z_N$ order
\begin{equation}
 K=
 \begin{pmatrix}
 0&N\\
 N&0
 \end{pmatrix},
 \qquad
 \Aanyon=\mathbb Z_N\times\mathbb Z_N,
 \label{eq:ZN_K}
\end{equation}
the electric and magnetic boundaries condense
\begin{equation}
 L_e=\langle(1,0)\rangle,
 \qquad
 L_m=\langle(0,1)\rangle.
 \label{eq:ZN_boundaries}
\end{equation}
Eq.~\eqref{eq:open_sector_dimension} gives
\begin{equation}
 \dim\cH_{ee}^{\rm top}=N,
 \qquad
 \dim\cH_{em}^{\rm top}=1.
 \label{eq:ZN_GSD}
\end{equation}
A junction between $e$- and $m$-type boundary segments carries a defect of
quantum dimension $\sqrt N$ and supports a generalized parafermion zero mode
~\cite{Lindner2012,ClarkeAliceaShtengel2013,BarkeshliJianQi2013}.  The $N=2$ case has quantum dimension $\sqrt2$
and is the analogue of a Majorana endpoint in an Abelian topological order.  This example shows
that non-Abelian endpoint degeneracy can arise even when the bulk anyons are
Abelian; it is a property of incompatible brane condensates.

\subsection{Endpoint motion and topological Berry phases}

Let an open Wilson line of type $a$ terminate on a movable endpoint.  If the
endpoint is adiabatically transported around a localized anyon $b$, the state
acquires the monodromy phase in Eq.~\eqref{eq:abelian_monodromy}.  Including
the electromagnetic charge $Q_a$, the endpoint Berry action has the schematic
form
\begin{equation}
 S_{\partial,\rm Berry}
 =\int\dd\tau\,
 \left[
 Q_a A_i(\bm X_a)\dot X_a^i
 +\cA_a^{\rm top}(q_a)\dot q_a
 \right].
 \label{eq:endpoint_Berry_action}
\end{equation}
The first term gives the ordinary Aharonov--Bohm phase; the holonomy of the
second gives the topological monodromy.  If $q_a$ is a single local coordinate,
a term $\cA(q_a)\dot q_a$ is locally a total derivative and contributes only
through global holonomy.  Local first-order dynamics requires a
multidimensional endpoint phase space or an additional conjugate variable.
For non-Abelian endpoints the topological connection becomes matrix valued,
but fusion-changing paths require a network and are deferred to the
splitting/joining theory.

\section{Moore--Read endpoints and Majorana junctions}
\label{sec:majorana}

\subsection{No local termination of a lone chiral Majorana}

The neutral sector of a Moore--Read interface contains a chiral Majorana field.
Such channels arise at Moore--Read boundaries and can also be isolated at
Moore--Read--331 interfaces; more complicated Pfaffian--anti-Pfaffian
interfaces may contain several neutral branches
\cite{Yang2017,ZhuShengYang2020}.
At an endpoint, fold every incident neutral channel onto the half-line
$x\geq0$ and separate incoming and outgoing fields.  With velocities absorbed
into the normalization,
\begin{equation}
 \widetilde\psi_\alpha
 =\sqrt{|v_\alpha|}\,\psi_\alpha,
 \label{eq:velocity_normalized_majorana}
\end{equation}
the boundary variation of the quadratic action is
\begin{equation}
 \delta S_\psi\big|_{x=0}
 =\frac{\ii}{2}\int\dd\tau\,
 \left(
 \widetilde{\bm\psi}_{\rm in}^T
 \delta\widetilde{\bm\psi}_{\rm in}
 -
 \widetilde{\bm\psi}_{\rm out}^T
 \delta\widetilde{\bm\psi}_{\rm out}
 \right).
 \label{eq:majorana_boundary_variation}
\end{equation}
A local, instantaneous, frequency-independent quadratic boundary condition
that cancels the variation and preserves the canonical anticommutators is
\begin{equation}
 \widetilde{\bm\psi}_{\rm out}
 =O\widetilde{\bm\psi}_{\rm in},
 \qquad O^T O=\mathbbm{1}.
 \label{eq:majorana_O_condition}
\end{equation}
The same orthogonality condition guarantees equality of incoming and outgoing
energy flux.  Within the free-Majorana channel description, the numbers of
incoming and outgoing real chiral modes must therefore match after all
anomaly-carrying host edges are included.  A dynamical endpoint may instead
produce a frequency-dependent unitary scattering matrix satisfying the
Majorana reality condition, but it cannot remove the net chiral central
charge.  A single Majorana arriving at an endpoint with no outgoing channel
has no local energy-conserving boundary condition.  The derivation is given in
Appendix~\ref{app:majorana_junction}.

A zero-dimensional Majorana operator cannot remove this obstruction by
itself.  It represents a finite parity degree of freedom only together with a
global partner and parity constraint, and it cannot carry a steady
one-dimensional thermal current. At a physical QH trijunction, the host-edge network required by the central-charge balance in Eq.~\eqref{eq:central_telescoping}
carries the compensating chiral flow.

\subsection{Localized zero modes and resonant absorption}

An endpoint defect may carry a localized Majorana $\gamma=\gamma^\dagger$.
The leading coupling to a continuing chiral channel is
\begin{equation}
 S_\gamma
 =\int\dd\tau\,
 \left[
 \frac{\ii}{2}\gamma\dot\gamma
 +\ii\lambda\gamma\psi(0)
 \right].
 \label{eq:gamma_coupling}
\end{equation}
The coupling is relevant and produces resonant absorption of the zero mode
into the edge.  For a single transmitting channel, the on-shell effect is a
unit-modulus phase
\begin{equation}
 S_\gamma(\omega)
 =\frac{\omega-\ii\Gamma}{\omega+\ii\Gamma},
 \qquad
 \Gamma>0,
 \label{eq:majorana_resonance_phase}
\end{equation}
where $\Gamma$ depends on the normalization of $\lambda$ and the edge
velocity.  The displayed ratio is given in the Fourier and chirality
convention used here; reversing either convention complex conjugates it.
There is no backscattering of a lone chiral channel, and the phase changes by
$\pi$ across the resonance.  The elementary scattering calculation is
included in Appendix~\ref{app:majorana_junction}.  The boundary renormalization-group
flow absorbs the free zero mode and reduces the boundary entropy according to
\begin{equation}
 S_{\rm UV}-S_{\rm IR}
 =\frac12\ln2
 \label{eq:majorana_entropy_drop}
\end{equation}
\cite{FendleyFisherNayak2009}.  Eq.~\eqref{eq:gamma_coupling} therefore
changes which endpoint boundary condition is realized, rather than providing
a sink for the gravitational anomaly.

\subsection{Ising branes and fusion-resolved open sectors}

After folding, the neutral endpoint problem is described by the nonchiral
Ising boundary conformal field theory.  Its elementary Cardy boundary
conditions are labeled by
\begin{equation}
 a\in\{1,\psi,\sigma\}.
 \label{eq:Ising_labels}
\end{equation}
The open-channel partition function is
\begin{equation}
 Z_{ab}(q)
 =\sum_cN_{ab}^{c}\chi_c(q),
 \label{eq:Cardy_partition}
\end{equation}
where $N_{ab}^{c}$ are the Ising fusion coefficients
\cite{Cardy1989,DiFrancesco1997}.  In particular,
\begin{align}
 Z_{11}&=\chi_1,
 &Z_{1\psi}&=\chi_\psi,
 &Z_{1\sigma}&=\chi_\sigma,
 \label{eq:Ising_Z1}
 \\
 Z_{\psi\psi}&=\chi_1,
 &Z_{\psi\sigma}&=\chi_\sigma,
 &Z_{\sigma\sigma}&=\chi_1+\chi_\psi.
 \label{eq:Ising_Z2}
\end{align}
Thus two $\sigma$-type endpoints give
\begin{equation}
 \cH_{\sigma\sigma}
 =\cV_1\oplus\cV_\psi,
 \label{eq:sigma_sigma_H}
\end{equation}
which is the fusion statement $\sigma\times\sigma=1+\psi$ realized as an
open-worldsheet Hilbert space.

The Cardy boundary entropies, or Affleck--Ludwig $g$ factors
\cite{AffleckLudwig1991}, are
\begin{equation}
 g_a=\frac{S_{a0}}{\sqrt{S_{00}}}
 =\frac{d_a}{\sqrt{\cD}},
 \label{eq:Cardy_g}
\end{equation}
with $\cD=2$, $d_1=d_\psi=1$, and $d_\sigma=\sqrt2$.  Hence
\begin{equation}
 g_1=g_\psi=\frac{1}{\sqrt2},
 \qquad
 g_\sigma=1.
 \label{eq:Ising_g_values}
\end{equation}
The relative factor $\sqrt2$ is the endpoint quantum dimension.  It agrees
with the entropy removed when a localized Majorana is absorbed by a chiral
edge.

A pair of well-separated $\sigma$ endpoints may be represented at low energy
by two Majorana operators $\gamma_L$ and $\gamma_R$, with splitting
\begin{equation}
 H_{\rm split}=\ii\epsilon\gamma_L\gamma_R.
 \label{eq:Majorana_splitting}
\end{equation}
For endpoints connected only through a gapped region, $\epsilon$ is
exponentially small.  If a gapless Majorana channel directly connects the
endpoints, the fusion sectors are generally split algebraically with system
size and are not an exponentially protected qubit.  The topological label
still controls the allowed conformal tower, but protection requires a fully
specified gapped network and global parity constraint.

\subsection{Worked trijunction: Moore--Read, 331, and vacuum}

A concrete QH example is obtained by taking phase $1$ to be the
Moore--Read state, phase $2$ to be the Halperin $331$ state, and phase $3$ to
be vacuum.  Their filling factors and chiral central charges are
$\nu_{\MR}=\nu_{331}=1/2~,~\nu_0=0$ and 
$c_{\MR}=\frac32 ~,~ c_{331}=2 ~,~ c_0=0$, respectively. Hence the three oriented interfaces obey Eqs.~\eqref{eq:hall_telescoping} and
\eqref{eq:central_telescoping} because
\begin{align}
 &\Delta\nu_{\MR,331}=0 ~,~
 \Delta\nu_{331,0}=\frac12 ~,~
 \Delta\nu_{0,\MR}=-\frac12,
 \label{eq:MR331vacuum_deltanu}
\\
 &\Delta c_{\MR,331}=-\frac12 ~,~
 \Delta c_{331,0}=2 ~,~
 \Delta c_{0,\MR}=-\frac32.
 \label{eq:MR331vacuum_deltac}
\end{align}

The mobile MR$|$331 segment has $\Delta\nu_{\rm MR,331}=0$, so the
electromagnetic Chern--Simons response does not by itself generate a charged
shape bracket.  At the charge-localized interface fixed point produced by
electron tunneling and sufficiently strong interfacial interactions, the
remaining propagating interface mode is a chiral Majorana fermion
~\cite{Yang2017}.  This is a property of that low-energy fixed point rather
than of an arbitrary microscopic MR$|$331 interface. Meanwhile, the outer 331$|$0 edge contains one charged chiral boson and one neutral $c=1$ channel, whereas the MR$|$0 edge contains one charged boson and one Majorana mode.  Together with the interface Majorana, these channels realize the telescoping Hall and central-charge differences in
Eqs.~\eqref{eq:hall_telescoping} and~\eqref{eq:central_telescoping}.
The sign of each chiral contribution is defined relative to the cyclic
orientation of its interface.  Anomaly balance constrains the complete set
of incident channels; it does not imply a one-to-one continuation of the
interface Majorana into the Majorana on the MR edge.

Therefore, a generic MR--331--vacuum trijunction is an interacting
boson--Majorana junction.  The orthogonal Majorana condition in
Eq.~\eqref{eq:majorana_O_condition} applies only when the relevant
neutral channels admit a quadratic Majorana description, for example at a
refermionizable or interaction-selected junction fixed point.  A
finite-dimensional endpoint defect alone still cannot terminate the
interface Majorana; the required chiral flow is carried by the full
host-edge network.

\section{Open-worldsheet spectra and endpoint spectroscopy}
\label{sec:spectrum}

\subsection{Directed-network quantization}

Because the interface modes are chiral, an open segment does not generically
support standing waves obtained by imposing independent reflecting conditions
at its two ends.  The spectrum belongs to the complete directed network made
from the mobile segment, host edges, and endpoint scattering matrices.  After
projecting each segment onto its propagating low-energy branches, let
$P_\alpha(\omega)$ be propagation along segment $\alpha$ and $S_a(\omega)$
the junction scattering matrix.  One circuit of the network defines
\begin{equation}
 \mathbb U(\omega)
 =\prod_{\rm circuit}P_\alpha(\omega)S_a(\omega).
 \label{eq:network_evolution}
\end{equation}
The eigenfrequencies obey
\begin{equation}
 \det\left[\mathbbm{1}-\mathbb U(\omega)\right]=0.
 \label{eq:network_quantization}
\end{equation}
For a single directed loop,
\begin{equation}
 \sum_\alpha k_\alpha(\omega)L_\alpha
 +\sum_a\delta_a(\omega)
 =2\pi(n+\alpha_0),
 \label{eq:single_loop_quantization}
\end{equation}
where $\alpha_0$ includes the bosonic winding or Majorana spin structure.  A
cubic shape segment contributes
\begin{equation}
 k_b(\omega)
 =\left(\frac{\omega}{\alpha_3}\right)^{1/3}
 +O(\omega),
 \label{eq:cubic_wave_number}
\end{equation}
while a Majorana segment contributes
$k_\psi(\omega)=\omega/v_\psi+\cdots$.  For a higher-spatial-derivative shape
equation, evanescent roots may also be required in a microscopic endpoint
matching problem; Eq.~\eqref{eq:single_loop_quantization} retains only the
propagating chiral root.  Endpoint phases therefore shift the finite-size
levels in a measurable way.  In particular, the resonant phase in
Eq.~\eqref{eq:majorana_resonance_phase} changes the Majorana level counting
across the zero-mode absorption crossover.

\subsection{Endpoint dynamics and resonances}

The endpoint coordinate $q_a$ may be dynamical, but a one-dimensional Berry
term $\cP_a(q_a)\dot q_a$ is locally a total derivative and cannot by itself
produce an endpoint resonance.  A local dynamical model must retain an
endpoint phase space.  For example, introduce a conjugate momentum $p_a$ and
write
\begin{equation}
 \begin{aligned}
 S_{q,a}
 ={}&\int\dd\tau\,
 \Bigl[
 p_a\dot q_a
 -\frac{p_a^2}{2m_a^{\rm end}}
 -\frac{k_a}{2}(q_a-q_a^{(0)})^2
 \Bigr]
 \\
 &+S_{\rm Berry}+S_{\rm coupling}.
 \end{aligned}
 \label{eq:endpoint_coordinate_action}
\end{equation}
Neither the endpoint mass $m_a^{\rm end}$ nor an equivalent dissipative
kernel is universal.  A genuinely first-order alternative requires at least a
two-dimensional endpoint phase space with nonzero Berry curvature.  Global
one-dimensional Berry holonomies may still shift the quantization condition
without generating local dynamics.

Linearizing Eqs.~\eqref{eq:end_force_balance}
and~\eqref{eq:end_moment_balance} couples $q_a$ to the endpoint values of the
shape field.  The resulting susceptibility probes the line tension, bending
rigidity, and the nonuniversal endpoint kinetic data.  Because the bulk shape
mode is chiral, an endpoint drive launches a directed wave rather than a
symmetric standing-wave pattern.

For a $\sigma$ endpoint, the same motion may modulate the zero-mode coupling
$\lambda(q_a)$ in Eq.~\eqref{eq:gamma_coupling}.  The endpoint susceptibility
then contains the Majorana resonance scale $\Gamma$.  Joint measurement of
the mechanical phase shift and the thermal or neutral response distinguishes
a purely geometric contact from a topological endpoint defect.

\subsection{Universal versus nonuniversal data}

The open-worldsheet observables separate into three classes.  The following
are topological or anomaly protected:
\begin{equation}
 \Delta\nu_{ij},
 \quad
 \Delta c_{ij},
 \quad
 \cS_{AB}=\Aanyon/(L_A+L_B),
 \quad
 N_{ab}^c,
 \quad
 d_a.
 \label{eq:universal_endpoint_data}
\end{equation}
The following geometric structures are fixed in form by the variational
principle:
\begin{equation}
 \bm F,
 \quad
 M,
 \quad
 \cA_{\rm op},
 \quad
 \text{Eqs.~\eqref{eq:end_force_balance} and \eqref{eq:end_moment_balance}}.
 \label{eq:geometric_endpoint_data}
\end{equation}
Their numerical values depend on the tension, bending rigidity, support
energy, and dividing-curve convention.  The contact energies, endpoint kinetic
kernels, charging capacitances, scattering matrices, zero-mode widths, and
curvature-dependent junction couplings are nonuniversal and must be matched
to microscopic calculations or experiment.

\section{Conclusion}
\label{sec:discussion}

We have developed an endpoint theory for a freely moving QH interface.  The
central point is that an open interface is a junction rather than a closed
chiral worldsheet with independently chosen boundary conditions.  The
embedding supplies mechanical force and moment, the charged sector carries a
boundary symplectic flux, and the incident topological phases determine which
charges and anomaly-carrying channels can continue through the endpoint.

The capped relative-area construction extends the material charge--shape
relation from a closed curve to an interval with endpoints sliding on support
curves.  It also separates the no-excess material sector from generic
endpoint charge transfer.  The latter requires an excess interfacial charge or
an explicit endpoint phase space and cannot be added while retaining the
strict material constraint unchanged.

After folding, the endpoint topological data are encoded by condensable
sectors together with any channels that remain ungapped.  For Abelian
nonchiral sectors, Lagrangian subgroups determine a basis of open Wilson-line
sectors.  For the Moore--Read neutral sector, a lone chiral Majorana cannot
terminate on a finite-dimensional defect.  A localized zero mode may change
the boundary condition, fusion sector, and boundary entropy, but the chiral
energy and gravitational anomaly must continue into an outgoing channel.

Several problems remain.  The endpoint kinetic terms and scattering matrices
must be derived from microscopic interface models, and geometric response
beyond the electromagnetic Chern--Simons sector may add momentum, spin, and
thermal boundary data.  Fermionic condensates require the corresponding spin
or super-modular refinement of the Abelian formulas.  Finally, processes in
which interfaces split, join, or reconnect change the worldsheet graph and
act on fusion space.  Their consistency involves $F$ and $R$ symbols and lies
beyond the fixed-topology open segment considered here.

\begin{acknowledgments}
The author is sincerely grateful to Professor Kun Yang for introducing him to
quantum Hall interface physics, and for the insightful discussions and
encouragement during his stay at the National High Magnetic Field Laboratory.
Although the author has since followed a path outside academia, the present
work continues those discussions.
\end{acknowledgments}

\appendix

\section{Boundary variation of the geometric energy}
\label{app:geometric_variation}

For completeness, start from
\begin{equation}
 \delta E_{0+2}
 =T_0\int\delta(\dd s)
 +T_2\int\left[2K\delta K\,\dd s+K^2\delta(\dd s)\right].
\end{equation}
Using Eq.~\eqref{eq:open_variation_identities},
\begin{equation}
\begin{aligned}
 \delta E_{0+2}
 ={}&\int\dd s\,
 \left[-T_0K+T_2(2K_{ss}+K^3)\right]\eta_n
 \\
 &+\left[
 T_0\eta_t+T_2(K^2\eta_t+2K\eta_{n,s}-2K_s\eta_n)
 \right]_L^R.
\end{aligned}
\end{equation}
Since $\delta\theta=\eta_{n,s}+K\eta_t$,
\begin{equation}
 K^2\eta_t+2K\eta_{n,s}
 =-K^2\eta_t+2K\delta\theta.
\end{equation}
The boundary term therefore becomes
\begin{equation}
 \left[
 (T_0-T_2K^2)\eta_t
 -2T_2K_s\eta_n
 +2T_2K\delta\theta
 \right]_L^R,
\end{equation}
which proves Eqs.~\eqref{eq:force_moment_form}
and~\eqref{eq:force_and_moment}.

\section{Capped area and interval transgression}
\label{app:open_area}

Let
\begin{equation}
 \cA_X
 =-\frac12\int_{\sigma_L}^{\sigma_R}\dd\sigma\,
 \eps_{ij}X^iX'^j.
\end{equation}
Integration by parts gives
\begin{equation}
 \delta\cA_X
 =\int\dd s\,\eta_n
 -\frac12\left[\eps_{ij}X^i\delta X^j\right]_L^R.
\end{equation}
At endpoint $a$,
\begin{equation}
 \delta X_a^j
 =\frac{\dd R_a^j}{\dd q_a}\delta q_a.
\end{equation}
Using Eq.~\eqref{eq:brane_area_primitive},
\begin{equation}
 \delta\cB_a
 =\frac12\eps_{ij}R_a^i\delta R_a^j.
\end{equation}
The $R$ endpoint term is canceled by $+\delta\cB_R$, while the $L$
endpoint term is canceled by $-\delta\cB_L$.  This proves
Eq.~\eqref{eq:open_area_variation}.  The same calculation with
$\delta\to\partial_\tau$ proves Eq.~\eqref{eq:open_area_time}.

For the interval transgression, integrate the magnetic two-form over the
parameter rectangle $(\sigma,r)$.  Stokes' theorem gives
\begin{equation}
 \int_{\sigma_L}^{\sigma_R}\dd\sigma\,\mathfrak a_\sigma
 =
 B\cA_{\rm op},
\end{equation}
where the two $r$-directed sides of the rectangle trace the support curves and
supply the cap terms.  At $\sigma=\sigma_a$ one may write
\begin{equation}
 \bm Y(\tau,\sigma_a,r)
 =
 \bm R_a[q_a(\tau,r)].
\end{equation}
Both $\partial_r\bm Y$ and $\partial_\tau\bm Y$ are then proportional to the
support tangent.  Their antisymmetric product vanishes, so
\begin{equation}
 \mathfrak a_\tau(\sigma_a)=0.
\end{equation}
Integrating Eq.~\eqref{eq:open_transgression_curvature} over the interval
therefore reproduces Eq.~\eqref{eq:open_area_time} without an additional
endpoint term.

\section{Open-sector counting for Lagrangian boundaries}
\label{app:open_counting}

The stretched Wilson-line label lies in $\Aanyon$.  Condensed endpoint anyons
act from the left by $L_A$ and from the right by $L_B$.  Since $\Aanyon$ is
Abelian, the double quotient reduces to
\begin{equation}
 L_A\backslash\Aanyon/L_B
 \simeq
 \Aanyon/(L_A+L_B).
\end{equation}
For finite Abelian groups,
\begin{equation}
 |L_A+L_B|
 =\frac{|L_A||L_B|}{|L_A\cap L_B|}.
\end{equation}
If $L_A$ and $L_B$ are Lagrangian, each has order
$\sqrt{|\Aanyon|}$.  Therefore
\begin{equation}
 \dim\mathbb C[\cS_{AB}]
 =
 \left|\frac{\Aanyon}{L_A+L_B}\right|
 =
 |L_A\cap L_B|,
\end{equation}
which proves Eq.~\eqref{eq:open_sector_dimension}.

\section{Majorana junction and localized-zero-mode scattering}
\label{app:majorana_junction}

For velocity-normalized Majoranas on half-lines, take
\begin{equation}
\begin{aligned}
 S
 =\frac{\ii}{2}\int\dd\tau\int_0^\infty\dd x\,
 \Bigl[&
 \widetilde{\bm\psi}_{\rm in}^T
 (\partial_\tau-\partial_x)\widetilde{\bm\psi}_{\rm in}
 \\
 &+
 \widetilde{\bm\psi}_{\rm out}^T
 (\partial_\tau+\partial_x)\widetilde{\bm\psi}_{\rm out}
 \Bigr].
\end{aligned}
\end{equation}
The boundary variation is Eq.~\eqref{eq:majorana_boundary_variation}, up to
the common convention for Grassmann endpoint variations.  Imposing
$\widetilde{\bm\psi}_{\rm out}=O\widetilde{\bm\psi}_{\rm in}$ gives
\begin{equation}
 \delta S\big|_{0}
 =\frac{\ii}{2}\int\dd\tau\,
 \widetilde{\bm\psi}_{\rm in}^T
 (\mathbbm{1}-O^T O)
 \delta\widetilde{\bm\psi}_{\rm in}.
\end{equation}
Thus $O^T O=\mathbbm{1}$.  The energy flux is proportional to
\begin{equation}
 \widetilde{\bm\psi}_{\rm in}^T\partial_\tau
 \widetilde{\bm\psi}_{\rm in}
 -
 \widetilde{\bm\psi}_{\rm out}^T\partial_\tau
 \widetilde{\bm\psi}_{\rm out},
\end{equation}
and vanishes for the same condition.

For the localized-zero-mode problem, unfold a single continuing channel onto
the full line and denote the fields immediately to the left and right of the
contact by $\psi_-$ and $\psi_+$.  With a symmetric point-contact
regularization, the equations following from Eq.~\eqref{eq:gamma_coupling}
may be written as
\begin{align}
 v(\psi_+-\psi_-)&=-2\lambda\gamma,
 \\
 \dot\gamma&=\lambda(\psi_++\psi_-).
\end{align}
For a mode of frequency $\omega$, let $\psi_+=S_\gamma(\omega)\psi_-$.  The
two equations give
\begin{equation}
 S_\gamma(\omega)
 =
 \frac{\omega-\ii\Gamma}{\omega+\ii\Gamma},
 \qquad
 \Gamma=\frac{2\lambda^2}{v},
\end{equation}
in the convention used in the main text.  Changes in the normalization of
$\gamma$, the contact prescription, or the Fourier convention rescale
$\Gamma$ or complex conjugate the displayed phase, but leave its unit modulus
and $\pi$ phase winding unchanged.

\end{document}